\documentstyle[aps,preprint,epsfig]{revtex}
\begin{document}
\def\ul{\underline}
\bibliographystyle{prsty}
\title{
Detection of gravitational waves
with quantum encryption technology
}
\author{Fabrizio Tamburini
\footnote{{\tt email:} fabrizio.tamburini@port.ac.uk},
Bruce A. Bassett \footnote{{\tt email:} bruce.bassett@port.ac.uk}
and Carlo Ungarelli\footnote{{\tt email:} carlo.ungarelli@port.ac.uk}}
\address{Relativity and Cosmology Group, School of 
Computer Science and Mathematics, Mercantile House,
University of Portsmouth, PO1 2EG, Portsmouth, UK}
\maketitle
\begin{abstract}
We propose a new technique for detecting gravitational waves using
Quantum Entangled STate (QUEST) technology. Gravitational waves reduce the
non-locality of correlated quanta controlled by Bell's
inequalities, distorting quantum encryption key statistics
away from a pure white noise.
Gravitational waves  therefore act as shadow eavesdroppers.
The resulting  colour distortions can, at least in principle, be
separated from noise and can differentiate
both deterministic and stochastic sources.
\vskip 2ex
\centerline{http://sun1.sms.port.ac.uk/cosmos/users/bruce/mousetrap/}
\vskip 2ex
\centerline{PACS numbers: 04.30.-w, 03.67Dd, 95.55Ym, 03.65Bz}

\end{abstract}

\vskip 18pc

\noindent Quantum cryptography provides a stunning
application of Einstein-Podolsky-Rosen (EPR)
correlations and Bell's inequalities
 \cite{bb84,BB92,ekert91,benn,bell,bou,jenn99}. Not only does
it promise perfectly
secure key distribution  but we argue that it may also  allow the
detection of the shadowy traces of gravitational waves whose
existence is the most important outstanding test
of Einstein's General Relativity and the subject of massive current
and next-generation experiments~\cite{LIGO,VIRGO,GEO,LISA}.

A general quantum encryptographic scheme consists of key generation
using entangled quantum states, by two parties (Alice and Bob) interested
in communicating securely. An attack may be made by an
eavesdropper (Eve) who secretly attempts to determine  the key as
the pairs of entangled EPR quanta travel to Alice and Bob. By performing
a sequence of measurements on these entangled pairs of photons,
Alice and Bob  determine the key they will use to encrypt their
message. The vital advantage quantum mechanics provides lies in
the {\em impossibility} that Eve
can intercept the secret key, made up of individual quanta,
without giving away her presence to Alice
and Bob, since such interception unavoidably  alters the entanglement
of the EPR pairs, as measured by violations of Bell's inequalities.

Variants of the standard BB84 protocol \cite{bb84,BB92}
based on the transmission of single pairs of EPR photons  have been used
recently in practical quantum key distribution over optic fiber networks
up to $48$ km in length \cite{losalamos}.  Similar experiments
\cite{jenn99,ADR,tittel98} have  illustrated the feasibility of quantum
encryption in practical situations
and the field is now sufficiently mature to be a tool in
fundamental research beyond the foundations of quantum mechanics.

In this letter we propose a simple {\em gedanken} experiment to detect
the effects of gravitational waves through the distortions they cause
in the statistics of the quantum keys determined by Alice and Bob.
The use of quantum encryption technology may be implemented in at
least two ways: one based on randomly
swapping polarisers, the second based on laser interferometry.

Consider the Ekert protocol\cite{ekert91} in
which Alice and Bob are equipped with randomly swapping polarizers.
Entangled pairs of photons are emitted in the singlet state
\begin{equation}
|\Psi \rangle =\frac 1{\sqrt{2}}[|H\rangle _A|V\rangle
_B+|V\rangle _A|H\rangle _B],
\end{equation}
where the photon $A$ is sent to Alice, and the photon $B$ to Bob.
$H$ and $V$ denote the horizontal and vertical polarizations,
prepared by a laser coupled to a 
parametric down-conversion device. The arrival time of the photons
at the polarizers is synchronized with their random swapping.

If a polarizer happens to be  correctly oriented, the incident photon
is detected,
and a ``1'' is recorded. Otherwise a ``0''. Repetition generates two
equal length  binary strings ${\bf A}$ and ${\bf B}$,
corresponding to the measurements of Alice's and Bob's detectors 
(see Fig. (1)).

Alice and Bob then publicly announce the orientations of
their polarizers corresponding to each element in ${\bf A}$ and ${\bf B}$.
They then eliminate the elements  of ${\bf A}$ and ${\bf B}$ corresponding
to non-coincident orientations of the two polarizers. The string entries
of the remaining subsets of ${\bf A}$ and ${\bf B}$  form the
two quantum keys, $K_{A}$ and $K_{B}$. In
the absence of gravitational waves and noise the two keys coincide,
$K_A = K_B$,  since  the  photon pairs were perfectly  entangled.

With the keys determined Alice proceeds with the transmission of a
message encrypted with her key, using e.g. logical
{\bf AND} or {\bf XOR}, to  Bob, who decodes it with his key, $K_B$
However, this is not of interest to us. Instead, cross-correlation
of the keys $K_A$ and $K_B$ allows, in principle, the detection of
gravitational waves.

This detection proceeds thanks to a fundamental property of quantum
cryptography: the keys derived from  an ideal experiment are
Markovian, pure white noise random strings of ``0''s and ``1''s.
The presence of a gravitational wave colours the cross-correlation statistics
so they are no longer white  (see Figures 1 and 2).

A gravitational wave introduces a discoloration by changing the
arrival time of the photons at Alice and Bob, by altering the detectors'
local time and the path length travelled by the photon. This implies
that, in the key strings $K_A$ and $K_B$, the
probability of a ``1'' (a  detection) is no longer equal to the
probability of a ``0'' (a non-detection). In addition
the two strings will no
longer coincide element by element: $K_A \neq K_B$.  In order to
analyse this effect one may construct the cross-correlation  
matrix between $K_A$ and $K_B$ (see Fig. 2) and search for
off-diagonal power.

Alternatively it is convenient to consider the string
${\bf K} \equiv K_A \otimes K_B$
formed using  an  appropriate operator $\otimes$, 
such as logical {\bf AND}. We then define the
{\it accumulated fluctuation} $\xi(t)$ as the
absolute value of the difference, for a given temporal length $t$,
of the number of non-detections, $N_{[0]}$, and detections,
$N_{[1]}$, in ${\bf K}$, {\em viz.} $\xi(t) \equiv |N_{[1]}-N_{[0]}|$.

$\xi(t)$  then obeys the  stochastic differential equation
\begin{equation}
\frac{d\,\xi}{d\,t}=\frac{\Gamma_{ph}}{2}(w(t)+h(t))\,,
\label{eqdiff}
\end{equation}
where $\Gamma_{ph}$ is the photons pair rate, $w(t)$ is a
stochastic process which describes the intrinsic noise of the
system and $h(t)$ is the strain produced by the gravitational
wave. The key feature of this equation is that, since the intrinsic
noise $w(t)$ is due only to the effective randomness of the
polarizer, it has the ideal statistical properties
\begin{equation}
\langle w(t)\rangle = 0 \,,\quad\quad
\langle w(t) w(t^{\prime}) \rangle = D\,\delta(t-t^{\prime}) \,\,.
\end{equation}
In the idealized case where complex and experiment-specific
noise sources (such as thermal and seismic fluctuations) 
are neglected, the intrinsic  fluctuations in
the time series $\xi$ extracted from the polarizers are  described by a
frequency-independent random process characterized by
the noise spectral density $D$ - the noise-induced mean square
fluctuations per unit frequency.

Since QUEST detectors use single photon pairs there is no shot
noise  and if $h(t) = 0$, the accumulated fluctuation is a
random process with zero mean and a linearly increasing
variance
\begin{equation}
(\mbox{Var}~ \xi)^2 = \langle\xi(t)\xi(t)\rangle=
\frac{\Gamma_{ph}^2}{4}\,
D\,t\,.
\end{equation}
A gravitational wave affects the statistical properties of the
accumulated fluctuations. For the sake of completeness, we
distinguish between (i) a deterministic gravitational wave (such as that
originating from a binary system of compact objects such as black holes 
or neutron stars)\cite{Flan} and (ii) a
stochastic background of gravitational waves (arising from inflation or
other cosmological sources) \cite{stoch1}. If a deterministic
gravitational wave is present the accumulated fluctuation is no
longer described by a zero-mean random variable. Rather its mean value is
\begin{equation}
\langle\xi(t)\rangle=\frac{\Gamma_{ph}}{2}\,\int^t_0\,d\tau h(\tau)\,.
\end{equation}
If instead a stochastic background  of gravitational waves
is present, the accumulated fluctuations are still described by a zero
mean variable, but the gravitational wave affects the
variance of $\xi$, which for a stationary gravitational wave background is 
\begin{equation}
(\mbox{Var}~\xi)^2=\frac{\Gamma_{ph}^2}{4}\,
D\,t\left[1+\frac{1}{D}\int^t_0\,d\tau H(\tau)\right]
\end{equation}
where $H(\tau)=\langle h(t)h(0)\rangle$.

In order to assess the sensitivity of our {\em gedanken} experiment
to a gravitational wave we need to estimate the noise background induced
by the effective randomness of the polarizer. Consider the data
set collected by one observer divided into sub-sets of $N$
points. Since each point corresponds to a photon,
$N=(\Gamma_{ph}/2)\tau_N$, where $\tau_N$ is the temporal length 
of the sub-set.
For each data sub-set, the background noise due
to the polarizer is then
\begin{eqnarray}
%&&
S_N=D/N\sim 2\cdot 10^{-43}\,\mbox{Hz}^{-1}\times
%\nonumber\\&&
\left(\frac{10^{8}\mbox{s}^{-1}}{\Gamma_{ph}}\right)
\left(\frac{\Theta_{sw}}{10^{-10}}\right)^2
\left(\frac{\tau_{coh}}{10^{-12}\mbox{s}}\right)
\left(\frac{10^{3}\mbox{s}}{\tau_N}\right)\,,
\label{Dcal}
\end{eqnarray}
where $\Theta_{sw}$ is  the percentage error
in the swapping of the random polarizers  and $\tau_{coh}$ is the
photon coherence time. At a given frequency $f$ and
for a data sub-set of about 20 minutes,
the characteristic amplitude of the noise
induced by the polarization swapper
$h_{rms}=(2f S_N)^{1/2}$ is then $\sim 6.3\times
10^{-22}(f/1\mbox{Hz})^{1/2}$. For comparison, the expected
characteristic noise amplitude for the LIGO interferometer around $200
Hz$ is $\sim 2\times 10^{-22}$. The noise associated with the
polarizers  is therefore minimized at frequencies lower than the
typical frequencies where large scale earth-based interferometers
reach their best sensitivity.

It is also important to stress that, unlike large scale
interferometers, this device effectively operates when 
the gravitational wavelength $\lambda_{GW}$ is
less than the distance $d_{AB}$ between the two receivers $A,B$.
In particular, the probability of ``detection'' and
``non-detection''  become equal when $\lambda_{GW} \gg 2d_{AB}$
and therefore there is a low frequency cut-off around $f \sim c/
2d_{AB}$. The precise response of such an experiment will depend
on seismic and thermal noise which in turn depend on the exact
experimental set-up. Since we are interested in general issues we do
not address this important issue here.

An alternative to the swapping polarizer set-up is to exploit a
quantum cryptographic protocol based on the continuous detection of photons
in interferometers,  relying on energy-time correlations  rather
than on polarization correlations \cite{tittel00}. We shall discuss in detail 
this implementation and noise-related issues in future work.

To clarify our proposal consider the effects of gravitational waves on
the famous Bell inequalities \cite{bell} describing quantum non-locality.
Both the polarizer and interferometer implementations can be unified in
the following formalism. Let $R_{ij}(\delta_A,\delta_B),~i,j = 0,1$ be
the number of time-correlated events detected by  Alice (A) and Bob (B)
as a function
of instrument parameters $\delta_i$, which will be polarization orientations
or phase shifts in the case of an interferometer setup.
The normalized  correlation coefficient of the measurements made  by the
detectors $A$ and $B$ is then ~\cite{ekert91}
\begin{eqnarray*}
&&E(\delta _A,\delta _B) =
%\\&&
\frac{R_{00}(\delta _A,\delta _B)-R_{01}(\delta
_A,\delta _B)-R_{10}(\delta _A,\delta _B)+R_{11}(\delta_A,\delta _B)}{
R_{00}(\delta _A,\delta _B)+R_{01}(\delta _A,\delta
_B)+R_{10}(\delta _A,\delta _B)+R_{11}(\delta _A,\delta _B)}\,.
\end{eqnarray*}
Following~\cite{CHSH}, one may then  define the composite operator
\begin{eqnarray}
S \equiv |E(\delta'_A,\delta''_B)+E(\delta'_A,\delta''_B)+E(\delta''{}_A,
\delta'_B)-E(\delta''{}_A,\delta''{}_B)|
\end{eqnarray}
where $\delta'_i$ and $\delta''_i$ represent specific values of the parameters
$\delta_i$. The quantity $S$ allows one to
test the degree of violation of Bell's inequalities; in particular,
the value $S = 2\sqrt{2}$ is achieved for maximal entanglement of the states.
Gravitational waves reduce $S$, as will a general eavesdropper.

We have outlined how quantum encryption technology may be
exploited to yield a potentially sensitive detector of weak cosmic
gravitational waves. These QUantum Entangled STate (QUEST)
detectors  are complementary to current and planned
interferometric detectors such as LIGO and LISA. While QUEST detectors
are not affected by shot noise,  detailed understanding
of all other relevant noise sources  is lacking and depends on exact
details of the detector set-up. Hence whether QUEST detectors can reach
the sensitivity of the interferometric detectors  by exploiting
the non-locality fundamental to quantum mechanics is still
unknown.   What is certain is that gravitational waves will act as
shadow eavesdroppers, reducing the  degree of entanglement between
quantum states controlled by Bell's inequalities, which is
precisely how they would be detected.

\acknowledgments
We thank Marco Bruni, David Kaiser, Philippos Papodopoulos
and Alberto Vecchio for insightful comments on the manuscript.

\newpage

\centerline{Captions}

{\bf Figure 1}

Schematic illustration of the proposed experiment: a source of
EPR photons repeatedly  sends a single pair of polarization-entangled
photons to
Alice and Bob who are equipped with randomly swapping
polarizers. A gravitational wave, by affecting the path length and local
proper time of the observer (here Bob) will reduce the probability of
detection of the photon causing distortions of the detection statistics
used to build the quantum keys (see Figure 2).

{\bf Figure 2}

The averaged cross-correlation matrix of sample $50$-element  keys $K_A$ and
$K_B$.  {\bf Inset:} The idealized white-noise case (without
gravitational waves). The diagonal
dominates in the large key length limit where the cross-correlation is simply
$\propto \delta_{ij}$.  {\bf The main figure} schematically shows
the effects of a deterministic gravitational wave which induces
off-diagonal  power representing non-white correlations.

\end{document}